\documentclass{PoS}

\usepackage[english]{babel}
\usepackage[utf8]{inputenc}
\usepackage[T1]{fontenc}

\definecolor{JGURed}{RGB}{193,0,42}

\title{Lattice calculations of the leading hadronic contribution to $(g-2)_\mu$}

\ShortTitle{Lattice calculations of the leading hadronic contribution to $(g-2)_\mu$}

\author{\speaker{Benjamin Jäger}, Hartmut Wittig\\
PRISMA Cluster of Excellence, Institut für Kernphysik and Helmholtz Institut Mainz,\\
Johannes Gutenberg-Universität, D-55099 Mainz, Germany\\
       E-mail: \email{jaeger@kph.uni-mainz.de}, \email{wittig@kph.uni-mainz.de}}

\author{Michele Della Morte \\
Instituto de Física Corpusculare\\
Universitat de València, E-46071 Valencia, España\\
E-mail: \email{dellamor@ific.uv.es}
}
    
\author{Andreas Jüttner\\
School of Physics and Astronomy\\
University of Southampton, Southampton, UK\\
E-mail: \email{a.juttner@soton.ac.uk}}

\abstract{\vspace{-11.5cm} \phantom{a} \hfill HIM-2012-4 \newline \phantom{a} \vspace{10.5cm} \newline 
We report on our ongoing project to calculate the leading hadronic contribution to the anomalous
magnetic moment of the muon $a_\mu^\mathrm{HLO}$ using two dynamical flavours of non-perturbatively
$\mathcal{O}(a)$ improved Wilson fermions. In this study, we changed the vacuum polarisation tensor 
to a combination of local and point-split currents which significantly reduces the numerical effort. Partially
twisted boundary conditions allow us to improve the momentum resolution of the vacuum polarisation tensor and 
therefore the determination of the leading hadronic contribution to $(g-2)_\mu$. We also extended the range of ensembles to
include a pion mass below $200\,\mathrm{MeV}$ which allows us to check the non-trivial chiral behaviour of $a_\mu^\mathrm{HLO}$.\newline
\begin{flushright} \includegraphics[width=3cm]{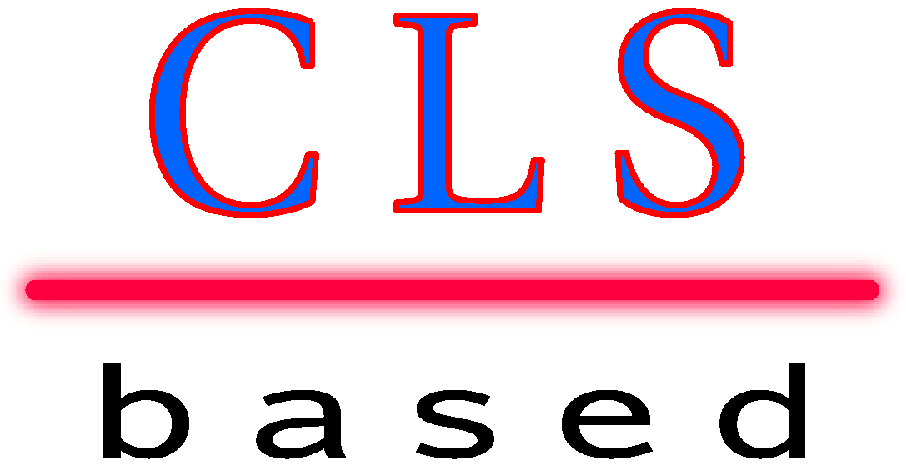} \end{flushright}
}

\FullConference{The 30th International Symposium on Lattice Field Theory\\
                 June 24 – 29,  2012\\
                 Cairns, Australia}

\begin{document}

\section{Introduction}
The anomalous magnetic moment of the muon $a_\mu = (g_\mu - 2)/2$ currently shows a persistent discrepancy of arround $3$ sigma
between experiment and theoretical predictions which may be a hint to physics beyond the Standard Model~[1,~2]. 
At the moment $a_\mu$ can be used to put exclusion limits on several BSM models, among many we mention the search for dark photons,
on which the experimental groups at the KPH Institute in Mainz are involved~[3]. The theoretical uncertainties 
of anomalous magnetic moment of the muon are dominated by hadronic contributions, from which the lowest order hadronic vacuum
polarisation $a_\mu^\mathrm{HLO}$ accounts for the largest uncertainty. At the moment the Standard Model prediction
of $a_\mu^\mathrm{HLO}$ is obtained by using the optical theorem to relate the $e^+e^- \rightarrow$ hadrons cross-section
data to the vacuum polarisation or by using the measured spectral functions from hadronic $\tau$ decays. 
We report on our ongoing project to determine the leading hadronic contribution to the anomalous
magnetic moment of the muon from first principles using Lattice QCD.

\section{Lattice Setup}
\begin{table}[h!]
\begin{center}
\begin{tabular}{ccccccc}
    	\hline
    	$\beta$ & $a$ $[\mathrm{fm}]$ & lattice & $L$ $[\mathrm{fm}]$ & 
    	$m_\pi$ $[\mathrm{MeV}]$ & $m_\pi L$ & Labels\\
    	\hline
    	$5.20$ & $0.079$ & $64 \times 32^3$ & $2.5$ & $473$, $363$, $312$ & 
    	$6.0$, $4.7$, $4.0$ & A3, A4, A5 \\
    	\hline
    	$5.30$ & $0.063$ & $64 \times 32^3$ & $2.0$ & $451$ & $4.7$ & E5 \\
    	$5.30$ & $0.063$ & $96 \times 48^3$ & $3.0$ & $324$, $277$ &  $5.0$, $4.2$
    	& F6, F7 \\
    	\textcolor{JGURed}{$5.30$} & \textcolor{JGURed}{$0.063$} &
    	\textcolor{JGURed}{$128 \times 64^3$} & \textcolor{JGURed}{$4.0$} &
    	\textcolor{JGURed}{$190$} & \textcolor{JGURed}{$4.0$} & \textcolor{JGURed}{
    	G8 } \\
    	\hline
    	$5.50$ & $0.050$ & $96 \times 48^3$ & $2.4$ & $430$, \textcolor{JGURed}{$330$} & $5.2$, \textcolor{JGURed}{$4.1$} & N4,
    	N5, \textcolor{JGURed}{N6} \\
    	\textcolor{JGURed}{$5.50$} & \textcolor{JGURed}{$0.050$} &
    	\textcolor{JGURed}{$128 \times 64^3$} & \textcolor{JGURed}{$3.2$} &
    	\textcolor{JGURed}{$260$} & \textcolor{JGURed}{$4.4$} & \textcolor{JGURed}{
    	O7 } \\
    	\hline
\end{tabular}
\caption{Summary of simulation parameters. The pion masses and the scale have been taken from~[13,~14,~15] and
are partly still preliminary.}
\label{tab}
\end{center}
\end{table}
We use non-perturbatively $\mathcal{O}(a)$ improved Wilson fermions with two dynamical flavours~[4]
and include a strange quark in a partially quenched framework. Our simulations were performed
on a subset of the gauge configurations generated within the CLS collaboration~[5]. The corresponding
simulation parameters are listed in table~1, where the ensembles highlighted in red have been
recently generated and added to the set analyzed in our earlier publication~[6]. Similar studies have been performed in the quenched approximation~[7,~8] and in
the theory with two~[9], three~[10,11] and four dynamical flavours~[12] .

\section{Vacuum Polarisation}
On the lattice in Euclidean space-time, the vacuum polarisation tensor is defined by a Fourier
transformation of a current-current correlator
\begin{equation}
\Pi_{\mu \nu}(q) = Z_V \, a^4 \sum_{x} e^{i q (x + a\hat{\mu}/2}) \left< J^\mathrm{c}_\mu(x) J^\mathrm{l}_\nu(0) \right> ,
\label{PImunu} 
\end{equation}
in which we use a combination of the local current $J^\mathrm{l}_\nu(x) =\bar{q}(x)\gamma_\mu q(x)$ and conserved point-split
current $J^\mathrm{c}_\mu(x)$ as done in~[11]. The conserved point-split current $J^\mathrm{c}_\mu(x)$ can be derived using Noether's
theorem and reads for Wilson fermions as
\begin{equation}
J^\mathrm{c}_\mu (x) = \frac{1}{2} \bigg(  \bar{q}(x+a\hat{\mu})
(1+\gamma_\mu) U_\mu^+(x) q(x) - \bar{q}(x) (1- \gamma_\mu) U_\mu(x)
q(x+a\hat{\mu}) \bigg).
\end{equation}
The Wick contraction of eq.~(3.1) leads to connected and disconnected contributions. In this work
we currently only consider the connected contributions, nevertheless two-flavour $\chi$PT can be used
to estimate the remaining disconnected piece to be approximately $-10\%$ of the connected one~[16]. The vacuum
polarisation $\Pi(q^2)$ can be extracted using the following relation:
\begin{equation}
\Pi_{\mu \nu}(q) =\left( \delta_{\mu\nu}
q^2 - q_\mu q_\nu \right) \Pi(q^2) .
\end{equation}
The use of the local current in the definition of the vacuum polarisation tensor requires a renormalization
of the operator. The relevant renormalization constant $Z_V$ was determined independently in~[17]. Using the combination of local and conserved currents is computationally cheaper.
Figure~1 shows a comparison of our previous work~[6] using two conserved point-split currents and the combination of local and conserved current
on the same gauge configurations. The unphysical constant $\Pi(0)$ turns out to be different in both determinations. Nevertheless
the subtracted vacuum polarisation which enters eq.~(3.4) agrees very well within the given statistical precision.
\begin{figure}[h!]
	\centering
	\hspace{-0.5cm}
	\begin{minipage}{0.64\linewidth}
	\centering
	\includegraphics[width=\linewidth]{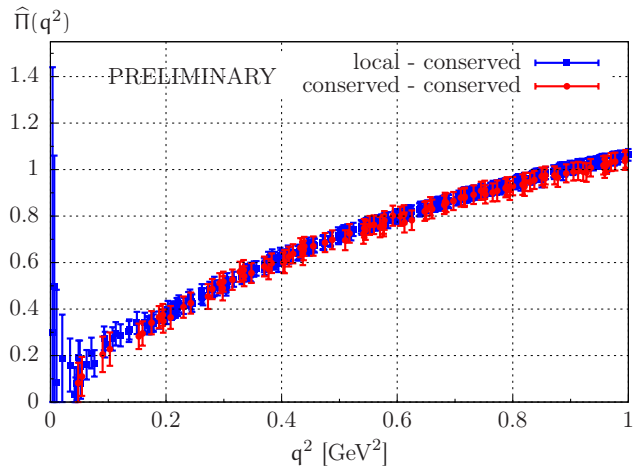}
	\end{minipage}
\caption{Comparison of the subtracted vacuum polarisation using two conserved point-split currents (red) and a
combination of local and conserved point-split current (blue). The subtracted vacuum polarisation is shown for the F6
ensemble ($\beta=5.3,$ $L=3.0\,\mathrm{fm},$ $m_\pi=324\,\mathrm{MeV}$).}
	\label{Bild1}
\end{figure}
The lowest order hadronic contribution to the anomalous magnetic moment of the muon $a_\mu^\mathrm{HLO}$ 
can be obtained by integrating
\begin{equation}
a_\mu = \left(\frac{\alpha}{\pi}\right) ^2 \int_0^\infty dq^2\,f(q^2)
\hat{\Pi}(q^2), \label{amu-int}
\end{equation}
which is the product of the subtracted vacuum polarisation $\hat{\Pi}(q^2) = 4 \pi \left(\Pi(q^2) - \Pi(0)  \right)$ 
and a function originating from QED~[18,~19]
\begin{equation}
f(q^2) = \frac{m_\mu^2 q^2 Z^3 (1- q^2 Z)}{1+m_\mu^2 q^2 Z^2}, \qquad\mathrm{with}\quad
Z=\frac{q^2- \sqrt{q^4 - 4 m_\mu^2 q^2}}{2 m_\mu^2 q^2}.
\end{equation}
Because of the shape of $f(q^2)$, the low momentum region of the vacuum polarisation gives the most
significant contribution to $a_\mu^\mathrm{HLO}$. To improve the momentum sampling we implemented twisted
boundary conditions to the valence quarks~[20]
\begin{equation}
q(x+L\,\hat{k}) = e^{i \theta_k} q(x).
\end{equation}
This modifies the momentum to $\frac{2\pi}{L} \vec{n} -\frac{\vec{\theta}}{L}$, where $\theta_k$ can be tuned to any real value. Using isospin symmetry,
the connected contribution of the vacuum polarisation can be understood as flavour non-diagonal and partially
twisted boundary conditions can be applied to evaluate it (see~[16]).

\section{Determination of $a_\mu^\mathrm{HLO}$}
\begin{figure}[h!]
\hspace{-0.2cm}
\begin{minipage}{0.64\linewidth}
	\centering
	\includegraphics[width=\linewidth]{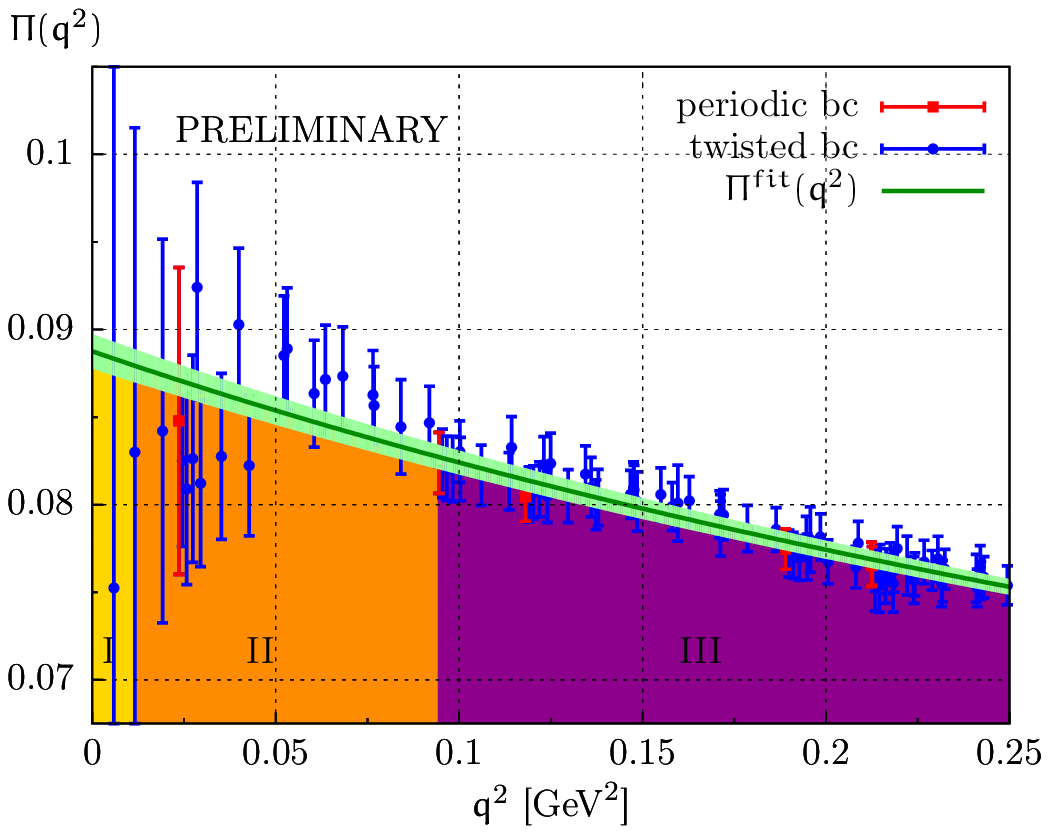}
\end{minipage}
\hspace{-0.6cm}
\begin{minipage}{0.40\linewidth}
	\centering
	\includegraphics[width=\linewidth]{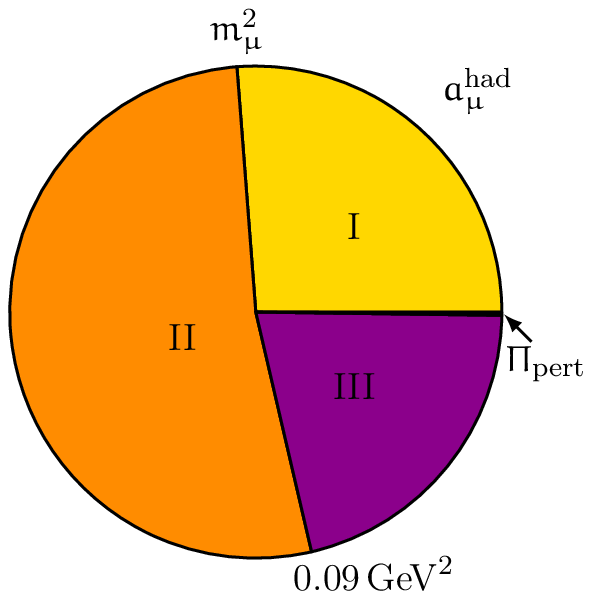}
\end{minipage}
\caption{{\bf Left:} The vacuum polarisation $\Pi(q^2)$ in the low-momentum region, computed on the G8 ensemble
 ($\beta=5.3,$ $L=4.0\,\mathrm{fm},$ $m_\pi=190\,\mathrm{MeV}$) 
using twisted and periodic boundary conditions. The green curve shows the result for a correlated
fit to a Pad\'e eq.~(4.4). {\bf Right:} The contribution to $a_\mu^\mathrm{HLO}$ from the integral in eq.~(3.4) separated by the different
momentum regions.}
\label{Bild2}
\end{figure}

The leading order hadronic contribution to $a_\mu$ can be obtained by integrating eq.~(3.4) numerically.
For this a continuous description of the vacuum polarisation $\Pi(q^2)$ is required. We perform
correlated least-square fits to the simulation data. Through varying the fit ansatz we are able to check for
systematic uncertainties, for which we choose
\begin{enumerate}
  \item[a)] a vector dominance model including a single vector
    \begin{equation}
  	\Pi(q^2)= a + \frac{b}{(q^2+c^2)} ,
  \end{equation}
  \item[b)] a vector dominance model with two vectors and one mass fixed to $m_V$ as proposed in~[11]
  \begin{equation}
  	\Pi(q^2)= a + \frac{b}{(q^2+c^2)} + \frac{d}{(q^2+m_V^2)}, 
  \end{equation}
   \item[c)] an ansatz motivated by a model for the hadronic cross section as used in~[8]
\begin{equation}
  	\Pi(q^2)= a + b\cdot\log(q^2+c^2) + \frac{d}{(q^2+e^2)},
  \end{equation}
    \item[d)] a model independent Pad\'e with various degrees as proposed in~[6,~21,~22], e.g.
  \begin{equation}
  	\Pi(q^2)= a + q^2\cdot \left(\frac{b}{(q^2+c^2)} + \frac{d}{(q^2+e^2)} \right).
  \end{equation}
\end{enumerate}
For the fit ansatz we imply that perturbation theory is matched smoothly at large momentum $q^2 >
2\,\mathrm{GeV}^2$, which allows us to reduce the number of fit parameters by one. Further details for the fitting
procedure and the matching to perturbation theory can be found in~[6]. All fits, except the vector
dominance model fit eq.~(4.1), agree within the statistical uncertainties. Figure~2 shows both the fit results for
the vacuum polarisation for small momentum and the clear improvement in the momentum resolution
achieved by partially twisted boundary conditions. The right panel
shows the individual contributions to $a_\mu^\mathrm{HLO}$ from different momentum ranges. It is important to note that without twisted
boundary conditions only a single data point lies in the momentum range covered by regions I and II. 
\section{Results}
\begin{figure}[h!]
\begin{minipage}{0.54\linewidth}
	\centering
	\includegraphics[width=\linewidth]{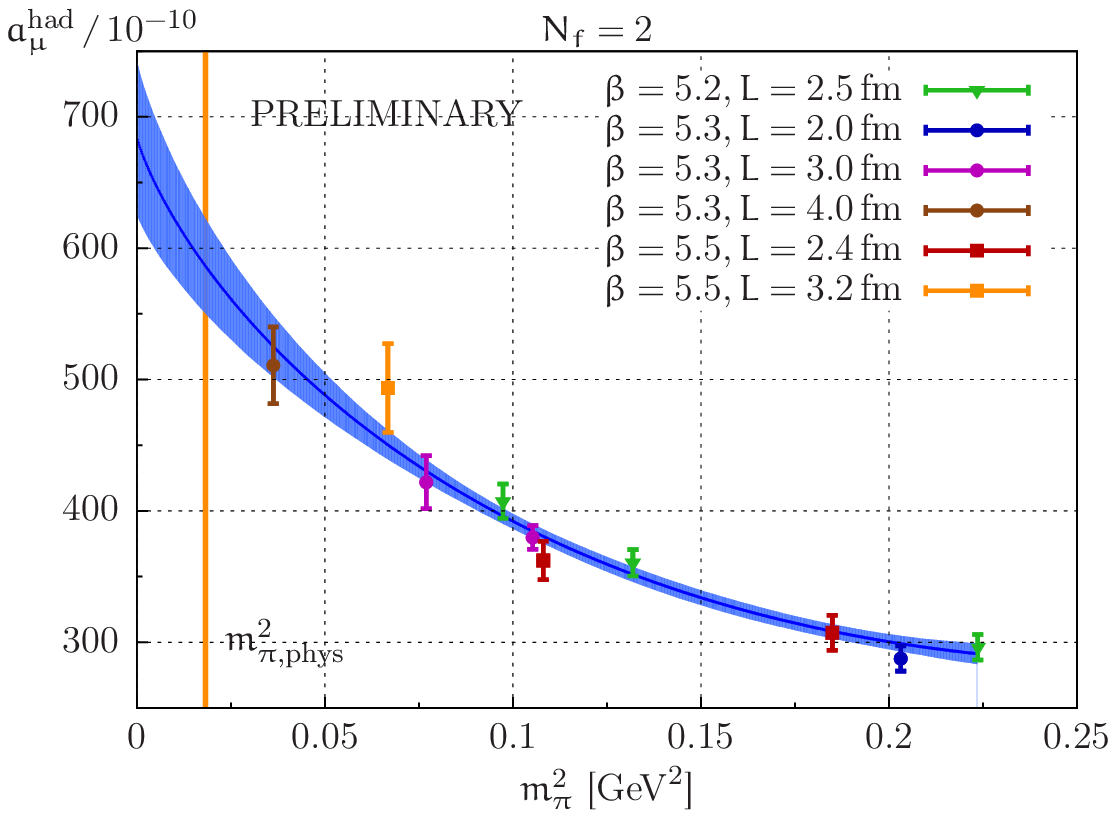}
\end{minipage}
\hspace{-0.8cm}
\begin{minipage}{0.54\linewidth}
	\centering
	\includegraphics[width=\linewidth]{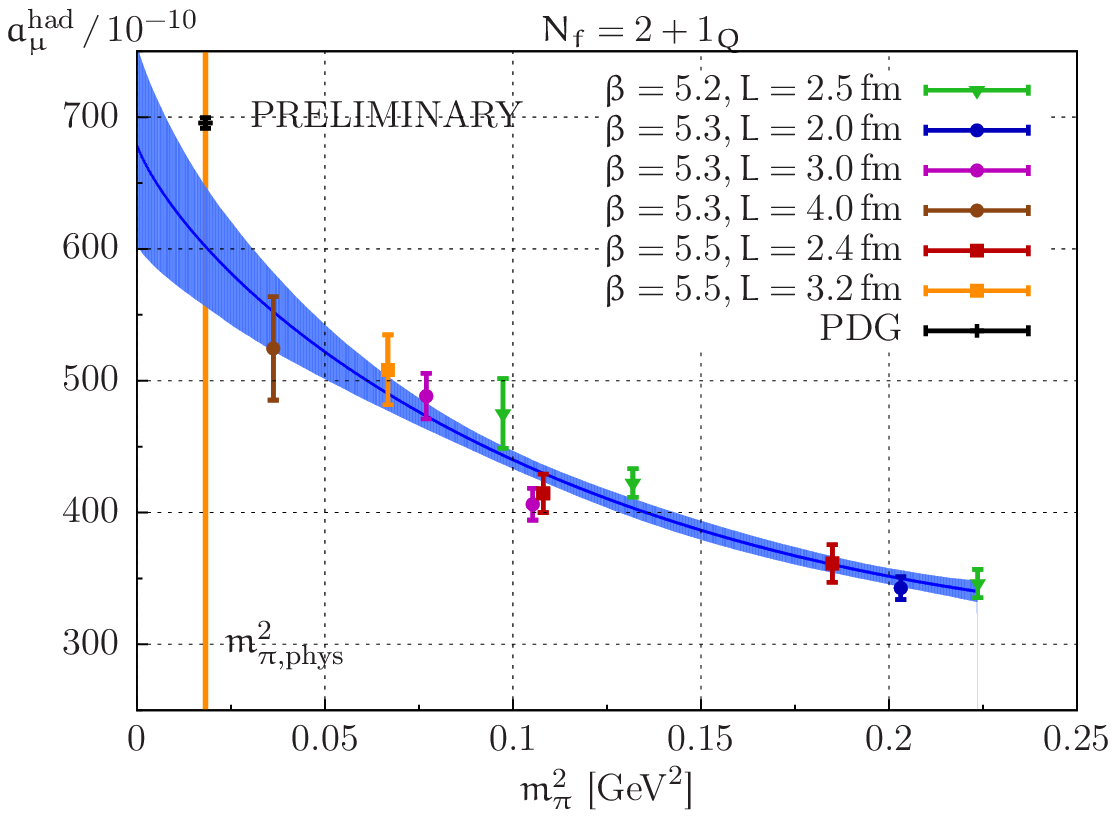}
\end{minipage}
\caption{{\bf Left:} The simulation results for $a_\mu^\mathrm{HLO}$ computed using two flavours, shown as function of $m_\pi^2$.
The chiral extrapolation (blue curve) is performed using an ansatz motivated by chiral perturbation theory. {\bf Right:} Equivalent results for $a_\mu^\mathrm{HLO}$
including a partially quenched strange quark.}
\label{Bild3}
\end{figure}
Repeating the fitting procedure for each of the ensembles listed in table~1 we obtain results for the
chiral behavior $a_\mu^\mathrm{HLO}$ shown in figure~3. The data show a clear non-linear behaviour and a rise
towards the physical pion mass. A chiral extrapolation is necessary to extract $a_\mu^\mathrm{HLO}$ at the physical
point. We use an uncorrelated fit to an ansatz motivated by chiral perturbation theory, i.e.
\begin{equation}
  	a_\mu^\mathrm{HLO}(m_\pi^2)= A + B \cdot m_\pi^2 + C \cdot m_\pi^2\, \log ( m_\pi^2). 
\end{equation}
In addition we perform a linear fit to the most chiral points 
\begin{equation}
  	a_\mu^\mathrm{HLO}(m_\pi^2)= A + B \cdot m_\pi^2
\end{equation}
to check for systematic uncertainties in the chiral extrapolation. Figure~3 shows the corresponding
fits as well as the uncertainties. We find that the $\chi$PT motivated fit describes the entire range
of data points quite well. With the current level of accuracy, the three flavour extrapolation misses the 
phenomenological value for $a_\mu^\mathrm{HLO}$, which might be caused by statistical downward 
fluctuations of the most chiral points. To verify the chiral behaviour we will increase our statistics on these ensembles. 

\section{Conclusions and Outlook}
The determination of the leading hadronic contribution to the anomalous magnetic moment
of the muon using lattice QCD has received considerable attention in the last years, but still requires additional improvements
to have an impact on phenomenology. The statistical uncertainty of the individual data points is 
currently of order $2-7\%$, where the largest uncertainties is obtained for the most chiral data points. 
Systematic uncertainties arising from chiral extrapolations and lattice artefacts
will come on top for the final lattice estimate. Partially twisted boundary conditions have proven to improve 
the sampling of the momentum for the vacuum polarisation and thereby reduce the statistical and systematic
uncertainties. The combination of conserved and local vector current reduces the numerical effort,
which allows one to improve the accuracy of the determination. Further details of our study will 
be published soon~[23]. Once all systematic uncertainties are under control, we will include a 
dynamical strange and charm quark into our simulations. By combining the strategy described here 
and "all mode averaging"~[24, 25] plus the approach in~[26] to directly estimate $\Pi(0)$, the connected
part of the leading order hadronic contribution to $(g-2)_\mu$ can be computed at level of less than few percent. 
At that point the dominating uncertainty will come from the missing disconnected contribution.
The application of the techniques in~[27] looks promising in this respect.

\phantom{TEST}

{\bf Acknowledgments:} We are grateful to our colleagues within the CLS project for sharing gauge
ensembles. Calculations of correlation functions were performed on the dedicated {Q}{C}{D} platform 
"Wilson" at the Institute for Nuclear Physics, University of Mainz. This work was granted access 
to the HPC resources of the Gauss Center for Supercomputing at Forschungzentrum Jülich, Germany,
made available within the Distributed European Computing Initiative by the PRACE-2IP, receiving
funding from the European Community’s Seventh Framework Programme
(FP7/2007-2013) under grant agreement RI-283493.28.


\begin{thebibliography}{99}
\bibitem{jegerlehner2009muon}
F.~Jegerlehner and A.~Nyffeler, 
{Physics Reports} {\bf 477} (2009) 1.

\bibitem{Benayoun:2011mm}
  M.~Benayoun, P.~David, L.~Del Buono and F.~Jegerlehner,
  Eur.\ Phys.\ J.\ C {\bf 72} (2012) 1848.

\bibitem{Merkel:2011ze}
H.~Merkel {\it et al.}  [A1 Collaboration],
Phys.\ Rev.\ Lett.\  {\bf 106} (2011) 251802.

\bibitem{jansen1998improvement}
K.~Jansen and R.~Sommer, 
{\bf B530} (1998) 185.

\bibitem{CLS}
https://twiki.cern.ch/twiki/bin/view/CLS/WebIntro (2012).

\bibitem{DellaMorte:2011aa}
  M.~Della Morte, B.~J\"ager, A.~J\"uttner and H.~Wittig,
  JHEP {\bf 1203} (2012) 055.

\bibitem{blum2003lattice}
T.~Blum, 
{Phys. Rev.} {\bf 91} (2003) 52001.

\bibitem{gockeler2004vacuum}
M.~G{\"o}ckeler, R.~Horsley, W.~K{\"u}rzinger, D.~Pleiter, P.E.L.~Rakow, and
  G.~Schierholz, 
  {Nucl. Phys.} {\bf B688} (2004) 135.

\bibitem{Feng:2011zk}
  X.~Feng, K.~Jansen, M.~Petschlies and D.~B.~Renner,
  Phys.\ Rev.\ Lett.\  {\bf 107} (2011) 081802.

\bibitem{Aubin:2006xv}
  C.~Aubin and T.~Blum,
  Phys.\ Rev.\ D {\bf 75} (2007) 114502.
  
\bibitem{Boyle:2011hu}
  P.~Boyle, L.~Del Debbio, E.~Kerrane and J.~Zanotti,
  Phys.\ Rev.\ D {\bf 85} (2012) 074504.
  
\bibitem{Hotzel}
G. Hotzel, {\em these proceedings}.  

\bibitem{Capitani:2011fg}
  S.~Capitani, M.~Della Morte, G.~von Hippel, B.~Knippschild and H.~Wittig,
  PoS LATTICE {\bf 2011} (2011) 145.

\bibitem{Capitani:2012gj}
  S.~Capitani, M.~Della Morte, G.~von Hippel, B.~J\"ager, A.~J\"uttner, B.~Knippschild, H.~B.~Meyer and H.~Wittig,
  Phys.\ Rev.\ D {\bf 86} (2012) 074502.

\bibitem{Georg}
G.~von Hippel, {\em private communication}.

\bibitem{della2010quark}
M.~Della~Morte and A.~J{\"u}ttner, 
{JHEP} {\bf 11} (2010) 154.

\bibitem{Della Morte:2005rd}
M.~Della Morte, R.~Hoffmann, F.~Knechtli, R.~Sommer and U.~Wolff,
JHEP {\bf 0507} (2005) 007.

\bibitem{Lautrup:1969fr}
  B.~E.~Lautrup and E.~De Rafael,
  Phys.\ Rev.\  {\bf 174} (1968) 1835.
  
\bibitem{Lautrup:1971jf}
  B.~E.~Lautrup, A.~Peterman and E.~De Rafael,
  Phys.\ Rept.\  {\bf 3} (1972) 193.

\bibitem{Bedaque:2004kc}
  P.~F.~Bedaque,
  Phys.\ Lett.\ B {\bf 593} (2004) 82.

\bibitem{Aubin:2012me}
  C.~Aubin, T.~Blum, M.~Golterman and S.~Peris,
  Phys.\ Rev.\ D {\bf 86} (2012) 054509.
  
\bibitem{Golterman}
M.~Golterman, {\em these proceedings}.
  
\bibitem{paper}
M.~Della Morte, B.~J{\"a}ger, A.~J{\"u}ttner and H.~Wittig, in preparation.
  
\bibitem{Blum:2012uh}
T.~Blum, T.~Izubuchi and E.~Shintani,
arXiv:1208.4349 [hep-lat].
  
\bibitem{Blum}
T.~Blum, {\em these proceedings}.

\bibitem{deDivitiis:2012vs}
G.~M.~de Divitiis, R.~Petronzio and N.~Tantalo,
arXiv:1208.5914 [hep-lat].

\bibitem{Vera}
V.~Gülpers, {\em these proceedings}.
\end{thebibliography}
\end{document}